\documentclass[aps,preprint,showpacs,preprintnumbers,amsmath,amssymb]{revtex4}
\usepackage{amsmath,mathrsfs,amsbsy,color,graphicx,bm,amsthm,amsfonts}
\usepackage{units}
\usepackage{bbm}
\usepackage{times}
\usepackage{dcolumn}
\usepackage{mathrsfs}
\usepackage{amsmath,amssymb,epsfig}
\usepackage{amsmath}
\newcommand{\udots}{\mathinner{\mskip1mu\raise1pt\vbox{\kern7pt\hbox{.}}
\mskip2mu\raise4pt\hbox{.}\mskip2mu\raise7pt\hbox{.}\mskip1mu}}
\begin{document}
\title{Reflecting boundary induced modulation of tripartite coherence harvesting }
\author{Shu-Min Wu$^1$\footnote{Email: smwu@lnnu.edu.cn},  Xiao-Ying Jiang$^1$, Xiang-Yue Yu$^1$,   Zhihong Liu$^2$ \footnote{Email: zhihongliu@hunnu.edu.cn (corresponding author)}, Xiao-Li Huang$^1$\footnote{Email: huangxiaoli1982@foxmail.com  (corresponding author)}}
\affiliation{$^1$  Department of Physics, Liaoning Normal University, Dalian 116029, China \\
$^2$  Department of Physics, Changzhi University, Changzhi, 046011, China
}


\begin{abstract}
We study the extraction of quantum coherence by three static Unruh-DeWitt (UDW) detectors that interact locally with a massless scalar vacuum field in the vicinity of an infinite perfectly reflecting boundary. Depending on the setup, the detectors are positioned either parallel or  orthogonal to the boundary, with their energy gaps chosen to satisfy the hierarchy $\Omega_C\geq \Omega_B\geq \Omega_A$. Our analysis reveals that decreasing the detector-boundary separation leads to a monotonic degradation of quantum coherence, whereas the same boundary effect can simultaneously preserve and even amplify the harvested quantum entanglement. Moreover, when the detectors possess distinct energy gaps, coherence extraction is further inhibited; strikingly, such non-identical configurations substantially enhance the efficiency of entanglement harvesting and markedly extend the range of detector separations over which non-negligible entanglement can be generated.
Nevertheless, the harvesting of nonlocal quantum coherence is achievable over a significantly broader range of detector separations than that of quantum entanglement.  Despite exhibiting similar overall behavior, orthogonal detector configurations outperform parallel ones in coherence harvesting, highlighting the quantitative influence of detector geometry.  Overall, our study reveals a hierarchical distinction between quantum coherence and entanglement as operational resources in structured vacuum fields: quantum coherence is not only more readily accessible across space but also more robust than entanglement, whereas entanglement exhibits richer features and can be selectively activated and enhanced through boundary effects and detector non-uniformity.
\end{abstract}

\vspace*{0.5cm}
\maketitle
\section{Introduction}
Quantum coherence stands as one of the central features of quantum mechanics, originates from the superposition principle of quantum states, and it constitutes the foundation of quantum interference phenomena. As a prerequisite for quantum entanglement and correlations, it also serves as a critical resource for quantum computing and information processing \cite{L1,L2,L3}, thus holding an indispensable position in physics and quantum information science. It exhibits significant application value in a range of cutting‑edge fields, including quantum metrology, quantum cryptography, energy transport in a biological system, and nanoscale thermodynamics \cite{L11,L12,L8,L9,L10,L7,LL7,L4,L5,L6}. Recently, research on quantum coherence has reached a relatively mature stage, and its framework has gradually expanded from single particle systems to multipartite systems, promoting the development of emerging directions such as incoherence teleportation and coherence localization \cite{L13,L14}. Quantification of quantum coherence was first systematically proposed by Baumgratz et al. \cite{L15} within the framework of resource theory, with the relative entropy of coherence (REC) and the $\mathrm{l_1}$-norm of coherence being two typical measures. Compared with the REC, the $\mathrm{l_1}$-norm of coherence is more easier in computation and analysis, and this simplicity makes it an essential tool for theoretical exploration and applications in quantum technologies.

Relativistic quantum information is fundamentally connected to the intrinsic correlations present in the quantum vacuum, which are themselves shaped by the underlying spacetime geometry \cite{SDF1,SDF2,SDF3,SDF4,SDF5,SDF6,SDF7,SDF8,SDF9,SDF10,SDF11,SDF12,SDF13,SDF14,SDF15,SDF16,SDF18,SDF19,SDF20,SDF21,SDF22,SDF23,SDF24,SDF25,SDF26,SDF27,SDF28,SDF29,SDF30,SDF31,SDF88}. The vacuum state of a quantum field exhibits not only local properties but also nonlocal correlations, which can be probed and extracted by
UDW detectors interacting locally with the field \cite{SDF32,SDF33,SDF34}. These harvested correlations can manifest as quantum entanglement, mutual information, or quantum discord between the detectors, and can even be distilled into Bell pairs, establishing the quantum vacuum as a practical resource for quantum information processing \cite{A1,A2,A3,A4,A5,A6,A7,A8,A9,A10,A11,A12,A13,A14,A15,A16,A17,A18,A19,A20}. This phenomenon, known as correlation harvesting, displays rich behaviors under nontrivial conditions such as reflecting boundary, curved spacetime backgrounds, or noninertial motion of the detectors. By coupling initially uncorrelated detector pairs to the vacuum, one can systematically extract both entanglement and broader quantum correlations, which is formalized as the correlation harvesting protocol.

In flat spacetime, the presence of a reflecting boundary modifies the fluctuations of quantum fields, and since the dynamical evolution of a detector system strongly depends on these fluctuations, such boundary-induced modifications have been shown to play a crucial role in controlling entanglement generation \cite{B1,B2,B3,B4}. Most previous studies have focused on the entanglement harvesting between two UDW detectors locally interacting with the fluctuating vacuum field \cite{A1,A2,A3,A4,A5,A6,A7,A8,A9,A10,A11,A12,A13,A14,A15,A16,A17,A18,A19,A20}, while investigations of tripartite entanglement harvesting remain relatively limited \cite{B5,B6,B7}. It is well recognized that nonlocal quantum coherence provides a more comprehensive measure of nonclassical correlations than entanglement alone. Motivated by this, one aim of the present study is to examine the harvesting of nonlocal quantum coherence using three UDW detectors in the presence of a reflecting boundary and to explore the internal relationships among tripartite coherence. Furthermore, we investigate the role of
non-identical detectors in this context, addressing whether introducing energy gap differences can enhance quantum coherence and extend the range of interdetector separations over which coherence can be effectively harvested.

Based on the above motivations, we present a comprehensive study of quantum coherence harvesting using three UDW detectors, labeled A, B, and C, positioned near a perfectly reflecting boundary. The detectors can be arranged either parallel or perpendicular to the boundary, and their energy gaps are assumed to follow a decreasing order: the energy gap of detector C is no smaller than that of detector B, while that of detector B is no smaller than that of detector A. We obtain several novel aspects of quantum coherence harvesting in the presence of a reflecting boundary: (i) we find that decreasing the distance between the detectors and the boundary generally suppresses quantum coherence in a monotonic fashion, while the same boundary can protect and even enhance the entanglement harvested between the detectors \cite{B1}. This indicates that boundary-induced modifications of the vacuum fluctuations have distinct impacts on different types of quantum resources, highlighting the contrasting sensitivities of coherence and entanglement to environmental structure; (ii) when the detectors are non-identical with differing energy gaps, the extraction of quantum coherence is further diminished. Surprisingly, these non-identical configurations significantly improve the efficiency of entanglement harvesting and expand the interdetector separation range over which a meaningful amount of entanglement can be generated. This suggests that detector asymmetry can serve as a practical tool to optimize entanglement extraction in structured quantum fields \cite{B8}; (iii) despite the inhibitory effects on coherence in certain scenarios, nonlocal quantum coherence remains harvestable across a considerably wider spatial range than entanglement; (iv) a comparison between parallel and orthogonal detector alignments shows that, despite similar qualitative trends, orthogonal-to-boundary configurations consistently yield greater harvested coherence, indicating that detector geometry quantitatively modulates coherence extraction. Taken together, these findings reveal a hierarchical structure among operational quantum resources: coherence is spatially more accessible and robust, whereas entanglement possesses richer structure and can be selectively amplified through boundary effects and detector heterogeneity. This distinction emphasizes the complementary roles of coherence and entanglement in relativistic quantum information processing and provides guidance for designing detector configurations to optimize specific quantum resources.

The paper is organized as follows. In Sec. II, we provide a brief review of the framework for the three UDW detector model. In Sec. III, we present a detailed analysis of how the reflecting boundary and the detectors' energy gaps affect the harvested tripartite coherence for both parallel and perpendicular configurations. Finally, Sec. IV summarizes our main findings and discusses their physical implications.

\section{ Vacuum tripartite coherence for UDW detector model}
The UDW detector model $D$ is a two-level quantum system with ground state $|0\rangle_D$ and excited state $|1\rangle_D$, separated by an energy gap $\Omega_D$. We consider three UDW detectors, labeled $A$, $B$, and $C$, each coupled locally to a massless quantum scalar field $\hat{\phi}(x, t)$. The interaction Hamiltonian is given by
\begin{eqnarray}\label{s1}
\hat{H}(t)=\sum_{D=A,B,C}\frac{d\tau_D}{dt}\lambda_D\chi_D(\tau_D(t))\hat{\mu}_D(\tau_D(t))\otimes\hat{\phi}[x_D(t)],
\end{eqnarray}
where $\tau_D$ denotes the proper time of detector $D$, $t$ is a common time parameter, $\lambda_D$ is the coupling strength, and $\chi_D(\tau_D)$ is the switching function \cite{B5,B6}. Here, $\hat{\phi}[x_D(t)]$ represents the pullback of the field operator along the trajectory of detector $D$. The model thus encompasses a broad parameter space, including three distinct switching functions, three coupling strengths, and three energy gaps \cite{A6,A7,A8}. In the interaction picture, the monopole moment operator of detector $D$ takes the form
\begin{eqnarray}\label{s2}
\hat{\mu}_D(\tau_D)=e^{i\Omega_D\tau_D}|1\rangle_D\langle0|_D+e^{-i\Omega_D\tau_D}|0\rangle_D\langle1|_D .
\end{eqnarray}

For simplicity, we assume a uniform coupling constant for all detectors, denoted as $\lambda := \lambda_D$, and that they are weakly coupled, satisfying $\lambda \ll 1$. The unitary operator $\hat{U}$, which governs the time evolution of the detectors and the field during the interaction, is generated by the interaction Hamiltonian in Eq.(\ref{s1}) and can be expanded via the Dyson series:
\begin{equation}\label{s3}
\begin{aligned}
\hat{U} &:= \mathcal{T} \exp\left(-i \int_{\mathbb{R}} dt \, \hat{H}(t)\right) = 1 + (-i\lambda) \int_{\mathbb{R}} dt \, \hat{H}(t) \\
&\quad + \frac{(-i\lambda)^2}{2} \int_{\mathbb{R}} dt \int_{\mathbb{R}} dt' \, \mathcal{T} \hat{H}(t) \hat{H}(t') + \mathcal{O}(\lambda^3),
\end{aligned}
\end{equation}
where $\mathcal{T}$ denotes the time-ordering operator.

We consider the detectors are initially prepared in the ground state $|0_{A} 0_{B} 0_{C}\rangle$ as $t\rightarrow -\infty$, while the field resides in a well-defined vacuum state $|0\rangle$. The initial density matrix can thus be expressed as $\rho_0 = |0_{A} 0_{B} 0_{C}\rangle \langle 0_{A} 0_{B} 0_{C}| \otimes |0\rangle \langle 0 |$. After the interaction, the final state of the detectors is obtained by tracing over the field degrees of freedom:
\begin{eqnarray}\label{s4}
\rho_{ABC} = \text{Tr}_{\phi}[ \hat{U} \rho_0 \hat{U}^{\dagger} ].
\end{eqnarray}
In the ordered basis $\{|0_A0_B0_C\rangle,|0_A0_B1_C\rangle,|0_A1_B0_C\rangle,|1_A0_B0_C\rangle,|0_A1_B1_C\rangle,|1_A0_B1_C\rangle,|1_A1_B0_C\rangle,\\
|1_A1_B1_C\rangle\}$, the density matrix $\rho_{ABC}$ can be shown to take the following general form, valid to all orders in the coupling:
\begin{eqnarray}\label{s5}
\rho_{ABC} =\left(\!\!\begin{array}{cccccccc}
r_{11} & 0 & 0 & 0 & r_{51}^{*} & r_{61}^{*} & r_{71}^{*} & 0 \\
0 & r_{22} & r_{32}^{*} & r_{42}^{*} & 0 & 0 & 0 & r_{82}^{*} \\
0 & r_{32} & r_{33} & r_{43}^{*} & 0 & 0 & 0 & r_{83}^{*} \\
0 & r_{42} & r_{43} & r_{44} & 0 & 0 & 0 & r_{84}^{*} \\
r_{51} & 0 & 0 & 0 & r_{55} & r_{65}^{*} & r_{75}^{*} & 0 \\
r_{61} & 0 & 0 & 0 & r_{65} & r_{66} & r_{76}^{*} & 0 \\
r_{71} & 0 & 0 & 0 & r_{75} & r_{76} & r_{77} & 0 \\
0 & r_{82} & r_{83} & r_{84} & 0 & 0 & 0 & r_{88}
\end{array}\!\!\right),
\end{eqnarray}
where the matrix elements $r_{ij}$ are determined by the detector parameters, such as the energy gaps and switching functions, as well as by their relative motion \cite{B5}.  To leading order in the coupling $\lambda$, the density matrix in Eq. (\ref{s5}) becomes:
\begin{eqnarray}\label{s6}
\rho_{ABC} =\left(\!\!\begin{array}{cccccccc}
1 - (P_A + P_B + P_C) & 0 & 0 & 0 & X_{BC}^* & X_{AC}^* & X_{AB}^* & 0 \\
0 & P_C & C_{BC}^* &  C_{AC}^* & 0 & 0 & 0 & 0  \\
0 & C_{BC} & P_B & C_{AB}^* & 0 & 0 & 0 & 0 \\
0 & C_{AC} & C_{AB} & P_A & 0 & 0 & 0 & 0 \\
X_{BC} & 0 & 0 & 0 & 0 & 0 & 0 & 0 \\
X_{AC} & 0 & 0 & 0 & 0 & 0 & 0 & 0 \\
X_{AB} & 0 & 0 & 0 & 0 & 0 & 0 & 0 \\
0 & 0 & 0 & 0 & 0 & 0 & 0 & 0
\end{array}\!\!\right)
+ \mathcal{O}\!\left(\lambda^4\right),
\end{eqnarray}
where
\begin{eqnarray}\label{s7}
P_D = \lambda^2 \int_\mathbb{R} d\tau_D \int_\mathbb{R} d\tau_D' \, \chi_D(\tau_D) \chi_D(\tau_D') e^{-i \Omega_ D(\tau_D - \tau_D')} W(x_D(\tau_D), x_D(\tau_D')),
\end{eqnarray}
\begin{eqnarray}\label{s8}
C_{DD'} = \lambda^2 \int_\mathbb{R} d\tau_D \int_\mathbb{R} d\tau_{D'}' \chi_D(\tau_D) \chi_{D'}(\tau_{D'}') e^{-i(\Omega_ D \tau_D - \Omega _{D'} \tau'_{D'})} W(x_D(\tau_D), x_{D'}(\tau'_{D'})).
\end{eqnarray}
The term $X_{DD'}$ is given by
\begin{eqnarray}\label{s9}
X_{DD'} &=& -\lambda^2 \int_\mathbb{R} d\tau_D \int_\mathbb{R} d\tau_{D'}' \, \chi_D(\tau_D) \chi_{D'}(\tau_{D'}') e^{i(\Omega_D \tau_D + \Omega_{D'} \tau'_{D'})} \nonumber \\
&& \times \Bigl[ \theta(t(\tau_D) - t(\tau'_{D'})) W(x_D(\tau_D), x_{D'}(\tau'_{D'})) \nonumber \\
&& + \theta(t(\tau'_{D'}) - t(\tau_D)) W(x_{D'}(\tau'_{D'}), x_D(\tau_D)) \Bigr],
\end{eqnarray}
and $W(x, x')$ denotes the vacuum Wightman function
\begin{eqnarray}\label{s10}
W(x, x') := \langle 0 | \phi(x) \phi(x') | 0 \rangle.
\end{eqnarray}
Furthermore, it is noted that $C_{D'D} = C^*_{D D'}$ with $D, D' \in \{A, B, C\}$ and $D \neq D'$. It is important to clarify that if $D = D'$, then $C_{DD} = P_D$ \cite{B6,B7}.
The term $P_D$ represents the transition probability of detector $D$: tracing $\rho_{ABC}$ in Eq.(\ref{s6}) over the Hilbert spaces of the other two detectors yields the reduced density matrix
\begin{eqnarray}\label{s11}
\rho_D = \left(\!\!\begin{array}{cccccccc}
1 - P_D & 0 \\
0 & P_D
\end{array}\!\!\right)
+ \mathcal{O}\!\left(\lambda^4\right).
\end{eqnarray}

A quantum state is considered coherent with respect to a complete set of states if it can be written as a nontrivial linear superposition of these states. This notion of quantum coherence stems directly from the superposition principle, which serves as a cornerstone of quantum theory. In this work, we study the harvesting of nonlocal coherence by three Unruh–DeWitt detectors, where the coherence cannot be attributed to any individual detector alone but instead reflects nonlocal correlations among them.  As quantitative measure of coherence, we employ the $\mathrm{l_1}$-norm of coherence \cite{L15}. For a system of dimension $n$ with a reference basis $\{|i\rangle\} _{i= 1, ...n}$, the $\mathrm{l_1}$-norm is defined as the sum of the absolute values of all off-diagonal elements of the density matrix $\rho$:
\begin{eqnarray}\label{s12}
C_{\mathrm{l_1}}(\rho_{ABC})=\sum_{i\neq j}|\rho_{i,j}|.
\end{eqnarray}
Applying this definition to the density matrix $\rho_{ABC}$ in Eq. (\ref{s6}) yields
\begin{eqnarray}\label{s13}
C_{\mathrm{l_1}}(\rho_{ABC}) = 2\{|X_{AB}|+|X_{BC}|+|X_{AC}|+|C_{AB}|+|C_{BC}|+|C_{AC}|\}.
\end{eqnarray}

\section{Tripartite coherence  harvesting for detectors aligned parallel and orthogonal to the
boundary}
In this section, we investigate how a reflecting boundary and differences in detector energy gaps influence  quantum coherence among three static UDW detectors. To clarify the role of the boundary, we introduce schematic diagrams illustrating the detector configurations. In particular, we analyze two representative geometries: detectors arranged parallel to the boundary and detectors arranged perpendicular to it.

\subsection{Tripartite coherence for the detectors aligned parallel to the boundary}
\begin{figure}
\begin{minipage}[t]{0.5\linewidth}
\centering
\includegraphics[width=3.0in,height=5.2cm]{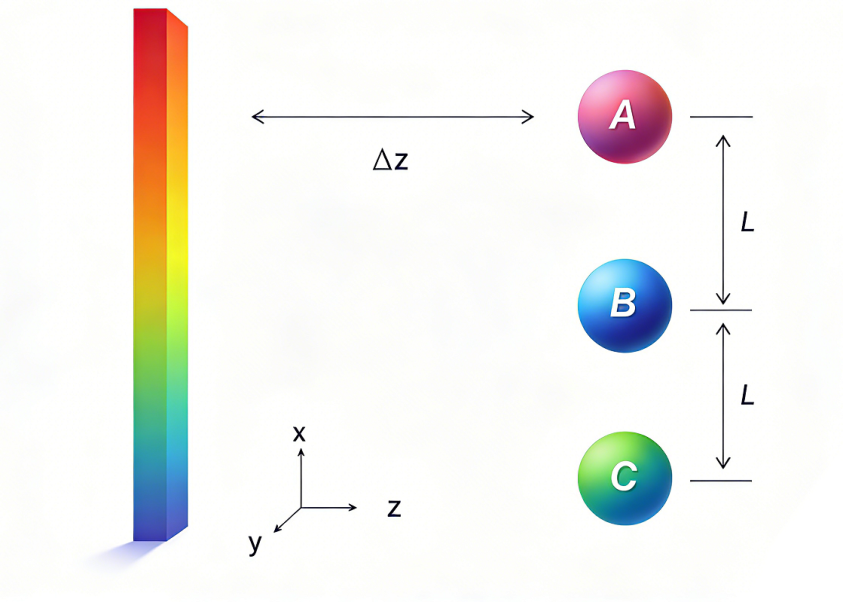}
\label{fig1a}
\end{minipage}%
\caption{ Schematic illustration of three static UDW detectors, designated as $A, B$ and $C$, aligned parallel to a reflecting boundary. Adjacent detectors are separated by a distance $L$ in the vertical direction and are positioned at a fixed distance $\Delta z$ from the boundary.}
\label{Fig.1}
\end{figure}

In this subsection, we consider three static detectors separated by an equal distance
$L$, each located at the same perpendicular distance
$\Delta z$ from the reflecting boundary. The detectors are aligned parallel to the boundary, as illustrated in Fig. \ref{Fig.1}. Under these conditions, the spacetime trajectories of the detectors are given by
\begin{eqnarray}\label{s14}
\nonumber &x_A:=\{t_A,x=0,y=0,z=\Delta z\},\quad x_B:=\{t_B,x=L,y=0,z=\Delta z\},\\
&x_C:=\{t_C,x=2L,y=0,z=\Delta z\}.
\end{eqnarray}

The transition probabilities $P_D$ and the correlation terms $C$ and $X$ can be determined using
Eqs.(\ref{s7})-(\ref{s9}), respectively. For these calculations, one must utilize the Wightman function for vacuum massless free scalar fields, which can be derived through the method of images

\begin{eqnarray}\label{s15}
 \nonumber W(x_D,x'_D)=-{\frac{1}{4\pi ^2}}\bigg[{\frac{1}{(t-t'-i\epsilon)^2-(x-x')^2-(y-y')^2-(z-z')^2}}\\
-{\frac{1}{(t-t'-i\epsilon)^2-(x-x')^2-(y-y')^2-(z+z')^2}}\bigg].
\end{eqnarray}
By substituting trajectory Eqs.(\ref{s14}) and (\ref{s15}) into Eq.(\ref{s7}), the transition probabilities can be obtained directly as
\begin{equation}\label{s16}
\begin{aligned}
 P_D=&{\frac{\lambda^2}{4\pi}}\bigg[e^{-\Omega_D ^2\sigma^2}-\sqrt{\pi}\Omega_D\sigma \rm{Erfc}(\Omega_D\sigma)\bigg]-{\frac{\lambda^2\sigma e^{-\Delta z^2/\sigma^2}}{8\sqrt{\pi }\Delta z}}\\
&\times\ \bigg\{\rm{Im}\bigg[e^{2i\Omega_D\Delta z}\rm{Erf}(i{\frac{\Delta z}{\sigma}}+\Omega_D\sigma)\bigg]-\sin(2\Omega_D \Delta z) \ \bigg\},
D\in \{A,B,C\},
\end{aligned}
\end{equation}
where $\rm{Erf(x)}$ denotes the error function and $\rm{Erfc(x)}:=1- \rm{Erf(x)}$. Similarly, the correlation terms $C_{AB}$ and $X_{AB}$ between detectors $A$ and $B$ in this context can also be evaluated as follows
\begin{eqnarray}\label{s17}
C_{AB}={\frac{\lambda^2}{4\sqrt{\pi}}}e^{-{\frac{(\Omega_B-\Omega _A)^2\sigma^2}{4}}}\big[f_{AB}(L)-f_{AB}(\sqrt{L^2+4\Delta z^2)}\big],
\end{eqnarray}
\begin{eqnarray}\label{s18}
X_{AB}=-{\frac{\lambda^2}{4\sqrt{\pi}}}e^{-{\frac{(\Omega_B+\Omega _A)^2\sigma^2}{4}}}\big[g_{AB}(L)-g_{AB}(\sqrt{L^2+4\Delta z^2)}\big],
\end{eqnarray}
with the auxiliary functions $f_{AB}(L)$ and $g_{AB}(L)$ defined as
\begin{eqnarray}\label{s19}
 \nonumber f_{AB}(L):=&{\frac{\sigma e^{-L^2/(4\sigma^2)}}{L}}\ \bigg\{\rm{Im}\bigg[e^{i(\Omega_A+\Omega_B)L/2}\rm{Erf}\bigg({\frac{iL+\Omega_B\sigma^2+\Omega_A\sigma^2}{2\sigma}}\bigg)\bigg]-\\
 &\sin\bigg[{\frac{(\Omega_A+\Omega_B)L}{2}}\bigg]\bigg\},
\end{eqnarray}
and
\begin{eqnarray}\label{s20}
\nonumber g_{AB}(L):=&{\frac{\sigma e^{-L^2/(4\sigma^2)}}{L}}\ \bigg\{\rm{Im}\bigg[e^{i(\Omega_B-\Omega_A)L/2}\rm{Erf}\bigg({\frac{iL+\Omega_B\sigma^2-\Omega_A\sigma^2}{2\sigma}}\bigg)\bigg]\\
 &+i\cos\bigg({\frac{(\Omega_B-\Omega_A)L}{2}}\bigg)\bigg\}.
\end{eqnarray}

Similarly, by replacing $\Omega_A$ with $\Omega_C$ in Eqs.(\ref{s16})-(\ref{s19}), we obtain $C_{BC}, X_{BC}, f_{BC}(L), g_{BC}(L)$. To derive $C_{AC}, X_{AC}, f_{AC}(L), g_{AC}(L)$, it is necessary to replace $\Omega_B$ with $\Omega_C$ in Eqs.(\ref{s16})-(\ref{s19}) as well as substituting $L$ with $2L$. Without loss of generality, we assume that the energy gap of detector $C$ is no smaller than that of detector $B$, and that the energy gap of detector $B$ is no smaller than that of detector $A$. Hence,  we can define $\Delta \Omega_{CB} := \Omega_C - \Omega_B \geq 0$ and $\Delta \Omega_{BA} := \Omega_B - \Omega_A \geq 0$. From Eqs.(\ref{s17}) and (\ref{s18}), it follows that the correlation terms $C$ and $X$ vanish when the duration parameter is sufficiently small compared with the energy‑gap difference, i.e., when $1/\sigma \ll \Delta \Omega$.

\begin{figure}
\begin{minipage}[t]{0.5\linewidth}
\centering
\includegraphics[width=3.0in,height=5.2cm]{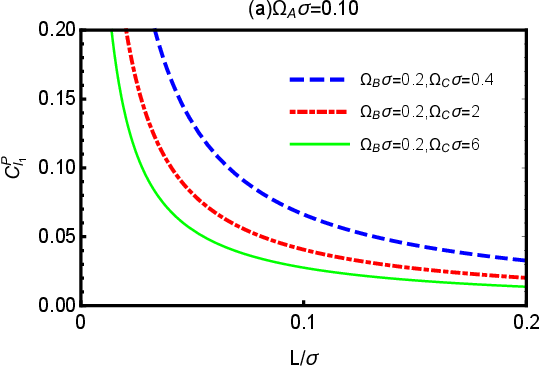}
\label{fig1a}
\end{minipage}%
\begin{minipage}[t]{0.5\linewidth}
\centering
\includegraphics[width=3.0in,height=5.2cm]{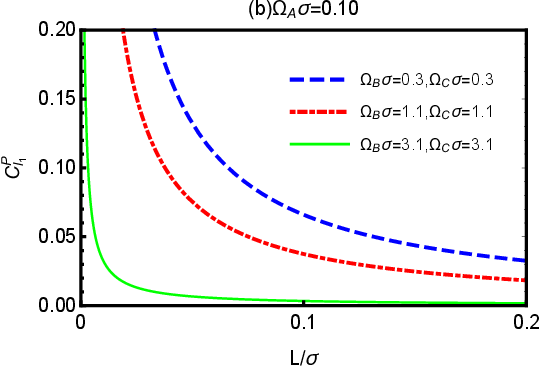}
\label{fig1b}
\end{minipage}%
\caption{Quantum coherence $C^{p}_\mathrm{l_1}$ as a function of detector separation $L/\sigma$, for $\Omega_A\sigma = 0.1$ and $\Delta z = 1$, with different values of $\Omega_B\sigma$ and $\Omega_C\sigma$.}
\label{Fig.2}
\end{figure}

To better understand how a reflective boundary affects the quantum coherence harvested by three detectors, we perform numerical calculations of the $\mathrm{l_1}$-norm of coherence. In this subsection, we designate the quantum coherence harvested by the three detectors arranged parallel to the boundary as $C^{p}_\mathrm{l_1}$.
Fig. \ref{Fig.2} shows the coherence $C^{p}_\mathrm{l_1}$ as a function of the detector separation $L/\sigma$ for several values of $\Omega_B\sigma$ and $\Omega_C\sigma$. A clear trend emerges:  the $\mathrm{l_1}$-norm of quantum coherence harvested between three UDW detectors decreases as their separation $L/\sigma$ increases. This suggests that placing detectors closer together can enhance coherence harvesting. When the sum of the energy gaps of the three UDW detectors is fixed, non-identical energy gaps for detectors
$B$ and $C$ give rise to a significantly larger amount of harvested quantum coherence compared to the identical-gap case.

\begin{figure}
\begin{minipage}[t]{0.5\linewidth}
\centering
\includegraphics[width=3.0in,height=5.2cm]{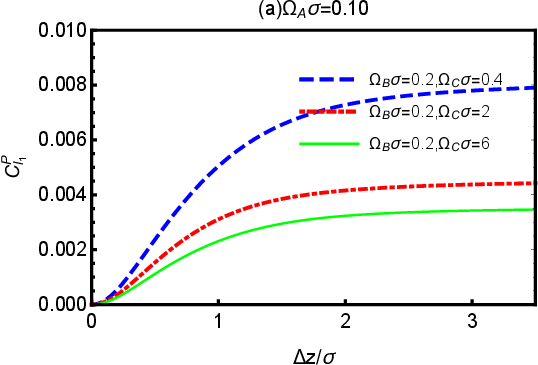}
\label{fig3a}
\end{minipage}%
\begin{minipage}[t]{0.5\linewidth}
\centering
\includegraphics[width=3.0in,height=5.2cm]{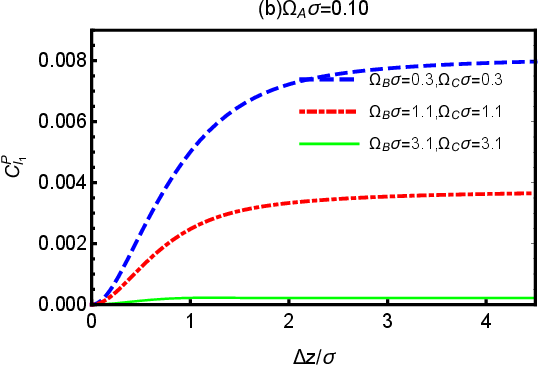}
\label{fig3b}
\end{minipage}%
\caption{Quantum coherence $C^{p}_\mathrm{l_1}$ as a function of the distance from the detectors to the boundary $\Delta z/\sigma$, for fixed $\Omega_A\sigma= 0.1$ and $L/\sigma = 1$, with varying values of $\Omega_B\sigma$ and $\Omega_C\sigma$.}
\label{Fig.3}
\end{figure}

In Fig. \ref{Fig.3}, we illustrate quantum coherence $C^{p}_\mathrm{l_1}$ as a function of the distance from the detectors to the boundary $\Delta z/\sigma$. As the detectors are moved away from the reflecting boundary, the harvested coherence grows smoothly from zero and saturates at a finite value in the large
$\Delta z/\sigma$ regime, signaling a progressive release from boundary-induced suppression. This monotonic trend demonstrates that the boundary acts as an effective constraint on coherence extraction at short distances. Interestingly, previous study has shown that the same boundary conditions may preserve or even amplify
detector-detector entanglement \cite{B1}. The contrast between these behaviors underscores that boundary-modified vacuum fluctuations influence distinct quantum resources in fundamentally different ways, revealing a clear resource-dependent response to spacetime structure.

\subsection{Tripartite coherence of detectors aligned orthogonally to the boundary}
\begin{figure}
\begin{minipage}[t]{0.5\linewidth}
\centering
\includegraphics[width=3.0in,height=5.2cm]{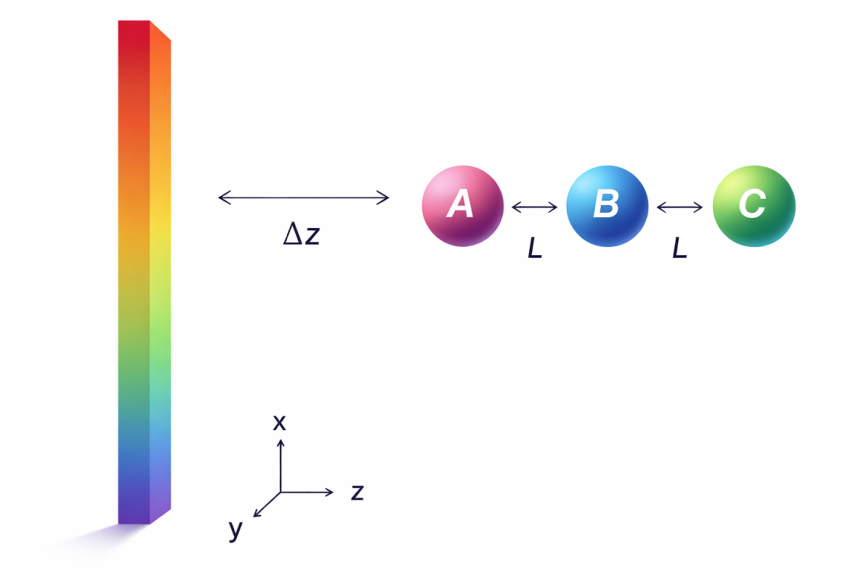}
\label{fig1a}
\end{minipage}%
\caption{Configuration of three static detectors aligned perpendicular to a reflecting boundary. Adjacent detectors are separated by a distance $L$, and the distance $\Delta z$ represents the separation between the boundary and the detector that is positioned closer to it.}
\label{Fig.4}
\end{figure}

It is well known that a reflecting boundary in flat spacetime breaks spatial isotropy. Consequently, the orientation of the detectors relative to the boundary can significantly influence quantum coherence harvested. In this subsection, we consider the configuration where the three detectors are aligned orthogonal to the boundary plane. As illustrated in Fig. \ref{Fig.4}, the spacetime trajectories of these static detectors are given by
\begin{eqnarray}\label{s21}
\nonumber &x_A:=\{t_A,x=0,y=0,z=\Delta z\}, \quad x_B:=\{t_B,x=0,y=0,z=L+\Delta z\}\\
&x_C:=\{t_C,x=0,y=0,z=2L+\Delta z\}.
\end{eqnarray}

\begin{figure}
\begin{minipage}[t]{0.5\linewidth}
\centering
\includegraphics[width=3.0in,height=5.2cm]{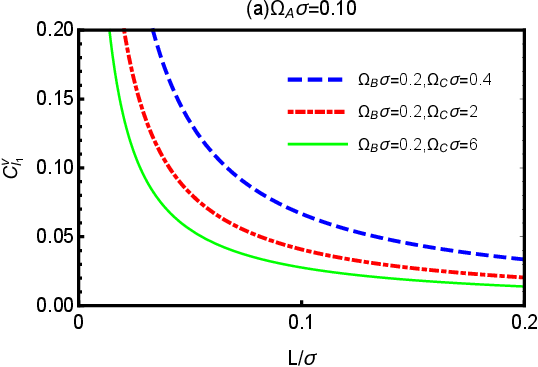}
\label{fig5a}
\end{minipage}%
\begin{minipage}[t]{0.5\linewidth}
\centering
\includegraphics[width=3.0in,height=5.2cm]{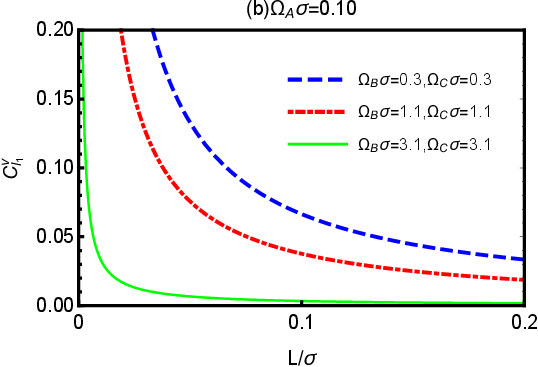}
\label{fig5b}
\end{minipage}%

\begin{minipage}[t]{0.5\linewidth}
\centering
\includegraphics[width=3.0in,height=5.2cm]{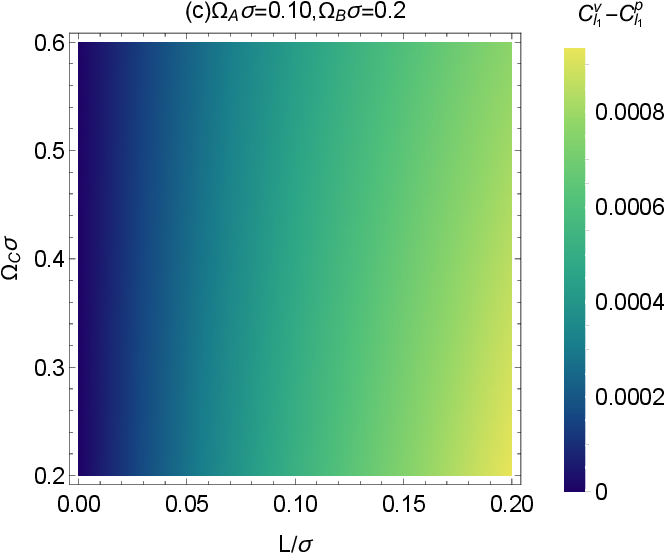}
\label{fig5c}
\end{minipage}%
\begin{minipage}[t]{0.5\linewidth}
\centering
\includegraphics[width=3.0in,height=5.2cm]{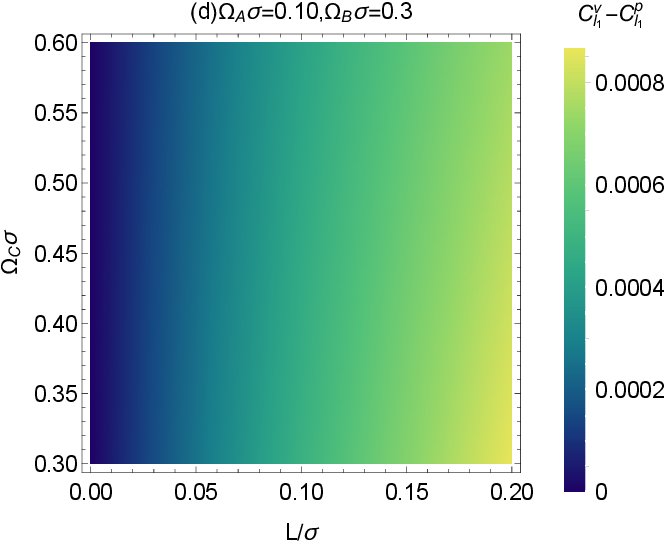}
\label{fig5d}
\end{minipage}%
\caption{Quantum coherence $C^{v}_\mathrm{l_1}$ (a) and (b) as a function of the distance between the detectors $L/\sigma$ for  different choices of $\Omega_B\sigma$ and $\Omega_C\sigma$. The difference $C^{v}_\mathrm{l_1}-C^{p}_\mathrm{l_1}$ (c) and (d) as  functions of the distance between the detectors $L/\sigma$ and  the energy gap $\Omega_C\sigma$. The  parameters are set to $\Omega_A\sigma = 0.1$ and  $\Delta z/\sigma = 1$.
}
\label{Fig.5}
\end{figure}

\begin{figure}
\begin{minipage}[t]{0.5\linewidth}
\centering
\includegraphics[width=3.0in,height=5.2cm]{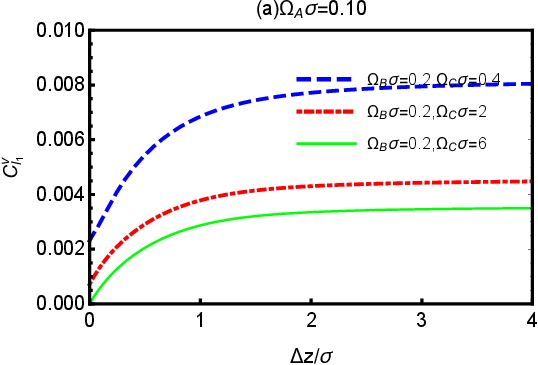}
\label{fig6a}
\end{minipage}%
\begin{minipage}[t]{0.5\linewidth}
\centering
\includegraphics[width=3.0in,height=5.2cm]{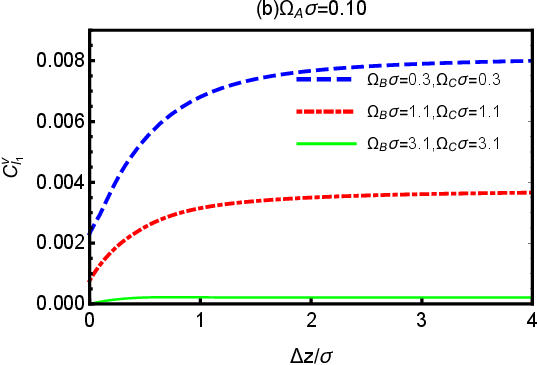}
\label{fig6b}
\end{minipage}%
\caption{Quantum coherence $C^{v}_\mathrm{l_1}$ as a function of the distance from the detectors to the boundary $\Delta z/\sigma$ for fixed $\Omega_A\sigma = 0.1$ and $L/\sigma = 1$, with different values of $\Omega_B\sigma$ and $\Omega_C\sigma$.}
\label{Fig.6}
\end{figure}

\begin{figure}
\begin{minipage}[t]{0.5\linewidth}
\centering
\includegraphics[width=3.0in,height=5.2cm]{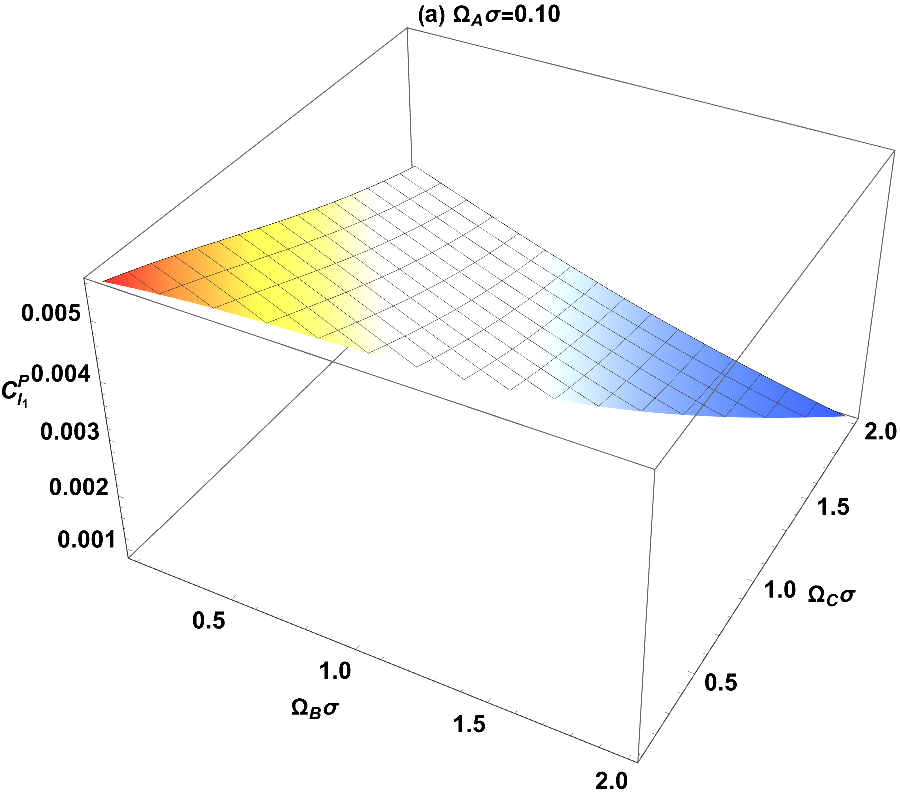}
\label{fig5a}
\end{minipage}%
\begin{minipage}[t]{0.5\linewidth}
\centering
\includegraphics[width=3.0in,height=5.2cm]{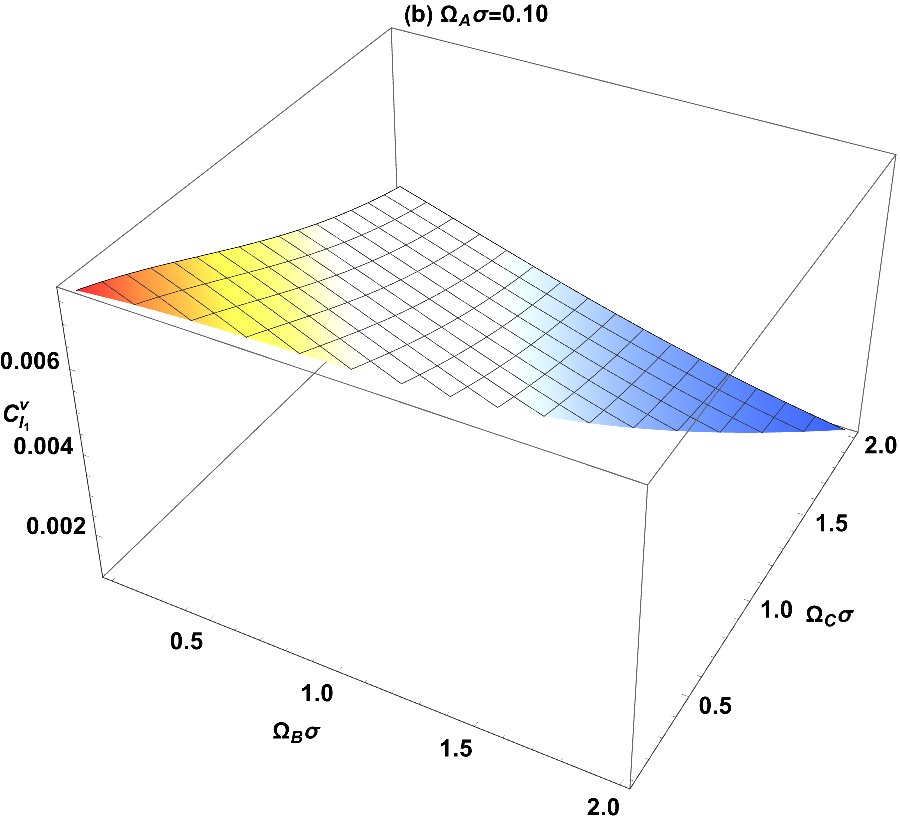}
\label{fig5b}
\end{minipage}%
\caption{Quantum coherence $C^{p}_\mathrm{l_1}$ and $C^{v}_\mathrm{l_1}$  as functions of the energy gaps $\Omega_B\sigma$ and $\Omega_C\sigma$ for fixed $\Omega_A\sigma= 0.1$, $L/\sigma = 1$, and $\Delta z/\sigma=1$.}
\label{Fig.7}
\end{figure}

The transition probability $P_A$ follows directly from Eqs.(\ref{s7}), (\ref{s15}), and (\ref{s21}), and is given by Eq.(\ref{s16}). The corresponding expressions for $P_B$ and $P_C$ are obtained by simply translating the coordinate $\Delta z$ in Eq.(\ref{s16}): replacing $\Delta z$ with $\Delta z + L$ yields $P_B$, and replacing it with $\Delta z + 2L$ yields $P_C$.
In the same way, the auxiliary equations $f _{AB}(L)$ and $g_{AB} (L)$ are derived from Eqs.(\ref{s19}) and (\ref{s20}) by accounting for the respective detector positions. Consequently, the correlation terms between detectors $A$ and $B$ are calculated as
\begin{eqnarray}\label{s22}
C_{AB}={\frac{\lambda^2}{4\sqrt{\pi}}}e^{-{\frac{(\Omega_B-\Omega _A)^2\sigma^2}{4}}}\big[f_{AB}(L)-f_{AB}(L+2\Delta z)\big],
\end{eqnarray}
\begin{eqnarray}\label{S30}
X_{AB}=-{\frac{\lambda^2}{4\sqrt{\pi}}}e^{-{\frac{(\Omega_B+\Omega _A)^2\sigma^2}{4}}}\big[g_{AB}(L)-g_{AB}(L+2\Delta z)\big].
\end{eqnarray}
Using the same method, the corresponding quantities for the other detector pairs $C_{BC}, X_{BC}, f_{BC}(L), g_{BC}(L)$, as well as $C_{AC}, X_{AC}, f_{AC}(L),$ and $g_{AC}(L)$, are obtained straightforwardly.
To examine how the boundary influences coherence harvesting differently depending on geometry, we compare the orthogonal configuration analyzed here with the parallel arrangement studied previously.
In this subsection, we denote quantum coherence harvested by the detectors arranged orthogonally to the boundary as $C^{v}_\mathrm{l_1}$, and the corresponding results are presented in Figs. \ref{Fig.5}-\ref{Fig.7}.

Fig. \ref{Fig.5}(a) and (b) display the harvested coherence $C^{v}_{\mathrm{l_1}}$ in the orthogonal configuration as a function of the detector separation $L/\sigma$. 
Fig. \ref{Fig.5}(c) and (d) explicitly show the difference $C^{v}_{\mathrm{l_1}}-C^{p}_{\mathrm{l_1}}$ as functions of the detector separation $L/\sigma$ and the energy gap $\Omega_C\sigma$.
A direct comparison with the parallel configuration shown in Fig. \ref{Fig.2} reveals a clear and consistent trend: for the same detector-boundary distance $\Delta z/\sigma = 1$ and identical spectral parameters, the orthogonal configuration systematically yields a larger amount of harvested coherence than the parallel one, as evidenced by the positive values of $C^{v}_{\mathrm{l_1}}-C^{p}_{\mathrm{l_1}}$ in Figs. \ref{Fig.5}(c) and (d).
This behavior demonstrates that orienting the detector array perpendicular to the reflecting boundary more effectively alleviates the boundary-induced suppression of quantum coherence under otherwise identical geometric and spectral conditions.

In Fig. \ref{Fig.6}, we present the behavior of quantum coherence $C^{v}_\mathrm{l_1}$ as a function of the normalized distance $\Delta z/\sigma$ between the detectors and the reflecting boundary. Similar to the case where the detectors are aligned parallel to the boundary, the coherence initially increases monotonically with $\Delta z/\sigma$ and gradually saturates at a steady value, indicating that the suppressive influence of the boundary diminishes with increasing separation. However, in contrast to the parallel configuration, $C^{v}_\mathrm{l_1}$ does not vanish at $\Delta z/\sigma = 0$ for the perpendicular alignment, and the overall coherence level is higher across the range. This can be attributed to the larger effective coupling distance between the detector system and the boundary in the perpendicular orientation, which weakens the boundary-induced suppression. As a result, stronger quantum coherence is achieved in the near-boundary region under this configuration. This finding provides not only a deeper insight into the role of boundary in modulating quantum coherence, but also a practical guideline for enhancing the stability and controllability of quantum properties in relativistic settings.

In Fig. \ref{Fig.7}, we show the harvested coherence as functions of the energy gaps $\Omega_B\sigma$ and $\Omega_C\sigma$, with $\Omega_A\sigma= 0.1$,  for detectors arranged both parallel and perpendicular to the reflecting boundary. The results indicate that the amount of harvested quantum coherence is maximized in the ideal configuration where all three detectors share identical energy gaps ($\Omega_A\sigma=\Omega_B\sigma=\Omega_C\sigma$). In contrast, when the detectors possess nonuniform energy gaps
($\Omega_A\sigma<\Omega_B\sigma<\Omega_C\sigma$), the extraction of quantum coherence is significantly suppressed. This behavior indicates that energy-gap asymmetry acts as an inhibiting factor for coherence harvesting. Interestingly, this trend is opposite to that observed for quantum entanglement, for which detector non-identicality has been shown to facilitate generation and enhancement  \cite{B8}. These results highlight a fundamental distinction in how different quantum resources respond to spectral inhomogeneity of the detectors.

In this work, we examine the $\mathrm{l_1}$-norm of quantum coherence by three non-identical Unruh-DeWitt detectors near a perfectly reflecting boundary. Our analysis reveals a fundamental structural property of tripartite coherence
\begin{eqnarray}\label{S31}
C^{p/v}_\mathrm{l_{1}}(\rho _{AB}) + C^{p/v}_\mathrm{l_{1}}(\rho _{BC}) + C^{p/v}_\mathrm{l_{1}}(\rho _{AC}) = C^{p/v}_\mathrm{l_{1}}(\rho _{ABC}),
\end{eqnarray}
where $C_{\mathrm{l_1}}(\rho_{AB}) = 2\{|X_{AB}|+|C_{AB}|\}$, $C_{\mathrm{l_1}}(\rho_{AC}) = 2\{|X_{AC}|+|C_{AC}|\}$, and $C_{\mathrm{l_1}}(\rho_{BC}) = 2\{|X_{BC}|+|C_{BC}|\}$.
Within the $\mathrm{l_1}$-norm framework and for the harvested detector state considered here, we find that the total tripartite coherence is exactly decomposable into pairwise contributions, leading to Eq. (\ref{S31}). This relation follows from the definition of coherence and the state structure, and implies the absence of genuinely global tripartite coherence in this setup.

\section{Conclusions}
In this work, we have performed a comprehensive analysis of quantum coherence harvesting using three static UDW detectors, labeled \(A\), \(B\), and \(C\), placed near a perfectly reflecting boundary. The detectors are arranged either parallel or perpendicular to the boundary, and their energy gaps satisfy the hierarchy \(\Omega_C \ge \Omega_B \ge \Omega_A\). Our study reveals several novel features of tripartite coherence extraction in structured vacuum fields: \textbf{(i) distinct boundary effects on coherence and entanglement:} the presence of the boundary affects coherence and entanglement differently. While decreasing the distance between the detectors and the boundary monotonically suppresses the harvested quantum coherence, the same boundary can preserve or even enhance entanglement between the detectors \cite{B1}. This demonstrates that
boundary-induced modifications of vacuum fluctuations exert distinct influences on different types of quantum resources; \textbf{(ii) impact of detector energy-gap asymmetry:}
the energy-gap asymmetry among the detectors has contrasting effects on coherence and entanglement. Non-identical detectors further reduce the harvested coherence, yet they substantially enhance entanglement generation and broaden the interdetector separation range over which significant entanglement can be achieved \cite{B8}; \textbf{(iii) robustness of nonlocal quantum coherence:}
despite the inhibitory effects on coherence under certain conditions, nonlocal quantum coherence can be harvested over a much wider spatial range than entanglement, highlighting its robustness as a quantum resource. This establishes a clear hierarchical distinction: coherence is more spatially accessible and resilient, while entanglement exhibits richer structure and can be selectively amplified through boundary effects and detector non-uniformity; \textbf{(iv) geometry-dependent coherence harvesting:}
a comparison between the parallel and orthogonal detector alignments reveals that, although the qualitative trends of coherence harvesting are similar, orthogonal-to-boundary configurations consistently yield higher amounts of harvested coherence than parallel-to-boundary setups. This indicates that detector geometry can quantitatively modulate coherence extraction; \textbf{(v) monogamy of tripartite coherence:}
we uncover a monogamy relation for tripartite coherence: the total coherence among the three detectors is fully accounted for by the sum of all bipartite coherences. In other words, no additional collective tripartite coherence global tripartite coherence can be harvested from the quantum vacuum.  Taken together, our findings illuminate the complementary roles of quantum coherence and entanglement in relativistic quantum information processing and provide guidance for designing detector arrangements and energy-gap configurations to optimize the harvesting of specific quantum resources.

\begin{acknowledgments}
This work is supported by the National Natural
Science Foundation of China (Grant Nos.12575056 and 12574384), Fundamental Research Program of Shanxi Province under Grant No.202503021212266, and Scientific and Technological Innovation Programs of Higher Education Institutions of Shanxi Province under Grant No.2025L120.	
\end{acknowledgments}


\end{document}